\documentclass[fleqn,preprint,showpacs,preprintnumbers,amsmath,amssymb]{revtex4}
\usepackage{graphicx}
\usepackage{dcolumn}
\usepackage{bm}

\begin{document}

\title{Comment on the article "Solitary waves and double layers in an ultra-relativistic degenerate dusty
electron-positron-ion plasma" [Phys. Plasmas \textbf{19}, 033705 (2012)]}
\author{M. Akbari-Moghanjoughi}
\affiliation{ Azarbaijan University of
Tarbiat Moallem, Faculty of Sciences,
Department of Physics, 51745-406, Tabriz, Iran}

\date{\today}
\begin{abstract}
More recently, N. Roy et al. [Phys. Plasmas \textbf{19}, 033705 (2012)] have investigated the occurrence of nonlinear solitary and double-layers in an ultrarelativistic dusty electron-positron-ion degenerate plasma using a Sagdeev potential method. They have considered a full parametric examination on Mach-number criteria for existence of such nonlinear excitations using the specific degeneracy limits of Chandrasekhar equation of state (EoS) for Fermi-Dirac plasmas. In this comment we point-out a misleading extension of polytropic EoS to study the Fermi-Dirac relativistically degenerate plasmas.
\end{abstract}


\maketitle

\section{Review and Comments}

It is known that the relativistic degeneracy is due to gigantic external pressure on stellar configuration caused by gravity. Such a change in the degeneracy state of fermions from the normal to relativistic one can alter the whole thermodynamic properties of dense matter \cite{kothary}. As a consequence the quantum degeneracy pressure of the electrons is altered. This change in the pressure is usually described by the polytropic shape $P_e\propto n_e^{\gamma}$ with $\gamma=4/3,5/3$ for the extreme cases of non- and ultra-relativistic degeneracies \cite{chandra1}.  It is obvious that such representation of equation of state is far from general and one can not extent such equation of state to the zero-temperature Fermi plasma of all density ranges. In a Fermi-Dirac plasma the degeneracy pressure is coupled with the electron density, $n_e$, or equivalently to the electron Fermi-momentum, $P_{Fe}$. Therefore, for a typical white-dwarf the density and electron degeneracy pressure increases as one move from outer layers towards the core. The exact form of Chandrasekhar pressure can be written as \cite{chandra2}
\begin{equation}\label{p}
{P_e} = P_0\left\{ {\eta\left( {2{\eta^2} - 3} \right)\sqrt {1 + {\eta^2}}  + 3\ln \left[ {\eta + \sqrt {1 + {\eta^2}} } \right]} \right\}{\rm{ }},
\end{equation}
in which $P_0={{\pi m_e^4{c^5}}}/{{3{h^3}}}$ and $\eta=(P_{Fe})/(m_e c)=(n_{e0}/n_{ec})^{1/3}$ ($n_{ec}\simeq 5.9 \times 10^{29} cm^{-3}$) is the relativity parameter with $P_{Fe}$ being the electron Fermi relativistic momentum. It is noted that the pressure is intimately related to the electron number-density of plasma and if one needs to define a general polytropic form of $P_e\propto n_e^{\gamma}$ for the pressure of degenerate electrons or positrons, one gets a parameter $\gamma=\ln (P_e/P_0)/\ln (n_{e0}/n_{ec})$ which is not independent of the equilibrium number-density of electrons. This fact has been ignored completely in Ref. \cite{roy}. It is also easy using the generalized Chandrasekhar relation Eq. (\ref{p}) to confirm mathematically that
\begin{equation}\label{pt}
{\gamma _0} = \mathop {\lim }\limits_{\eta  \to 0} \gamma  = 5/3,\hspace{3mm}{\gamma _\infty } = \mathop {\lim }\limits_{\eta  \to \infty } \gamma  = 4/3,
\end{equation}
as expected. Figure 1. shows the complicated dependence of the parameter, $\gamma$, on the electron equilibrium number-density. It is clearly observed that, the use of such polytropic form for general description of Fermi-Dirac plasmas is useless, unless, the complex relation of the parameter, $\gamma$, and the electron number-density has to be taken into account in formulation, which is obviously absent in Ref. \cite{roy}. It is further remarked that in this model the value of $\gamma$, hence the pressure, diverges for the typical white-dwarf electron-density of $n_{e0}\simeq 5.9 \times 10^{29} cm^{-3}$, while, for the exact pressure one gets a finite value. Thus, the generalized form of Chandrasekhar EoS should be used in order to evaluate the possibility of nonlinear excitations for the whole range of the relativity parameter (electron number-density) as in Refs. \cite{akbari, masood}

\newpage

\textbf{FIGURE CAPTIONS}

\bigskip

Figure-1

\bigskip

The variation of parameter, $\gamma$, in the artificial polytropic model for electron degeneracy pressure, $P_e\propto n_e^{\gamma}$, in white-dwarfs. The dashed lines represent the extreme degeneracy limits $\gamma=4/3$ and $\gamma=5/3$.

\begin{figure}[ptb]\label{Figure1}
\includegraphics[scale=.85]{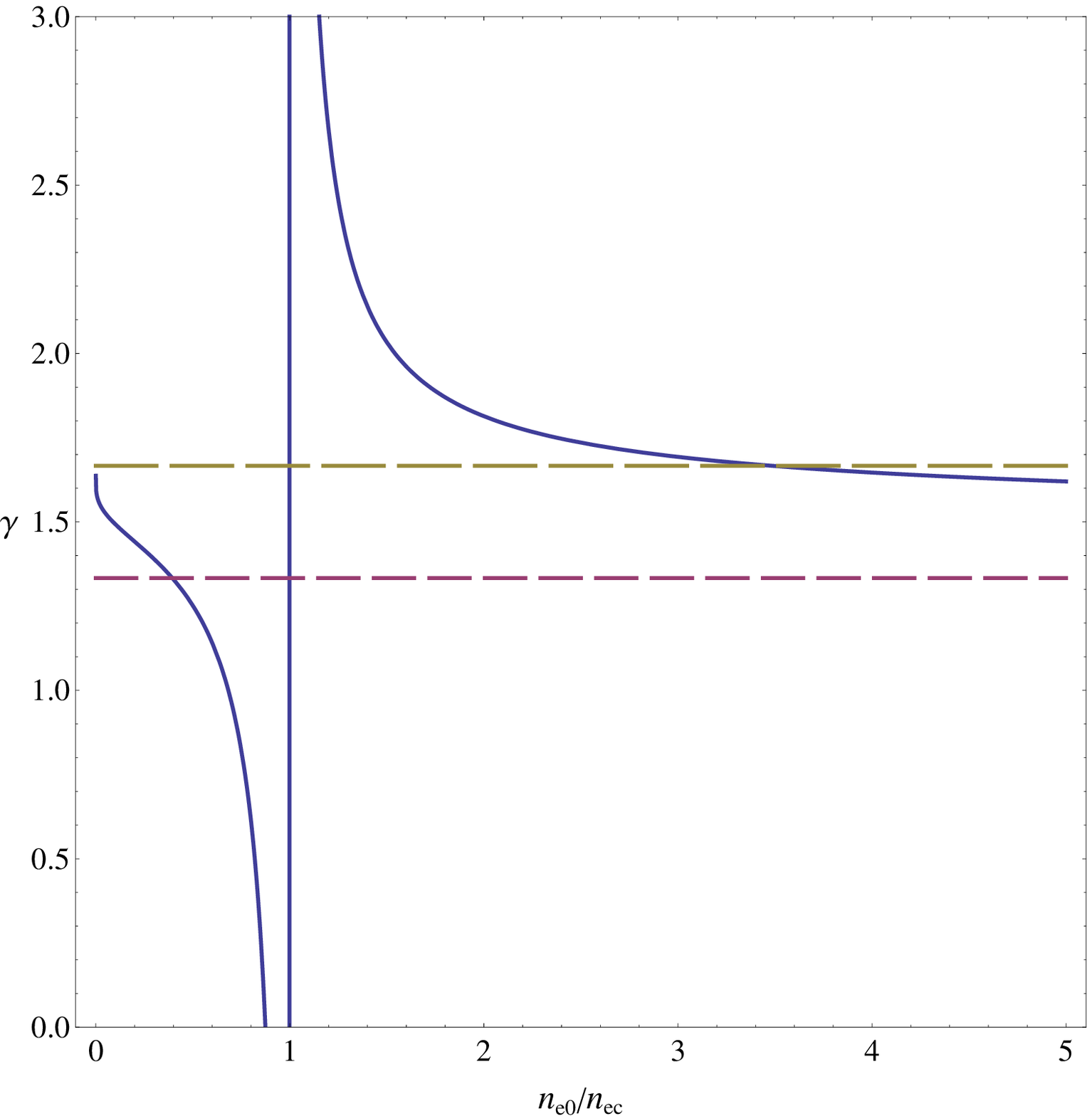}\caption{}
\end{figure}

\end{document}